\documentclass[11pt]{article}
\usepackage[T1]{fontenc}
\usepackage{mathtools}
\usepackage{amssymb}
\usepackage{dsfont}
\usepackage{slashed}
\usepackage{fullpage}
\usepackage{cite}
\usepackage{currvita}
\usepackage[colorlinks=true,linkcolor=blue,citecolor=magenta,linktocpage=true]{hyperref}
\usepackage{braket}
\usepackage{titlesec}
\titleformat*{\section}{\normalsize\bfseries}
\titleformat*{\subsection}{\normalsize\bfseries}
\titleformat*{\subsubsection}{\normalsize\bfseries}
\makeatletter
\renewcommand{\@dotsep}{1000}

\def\be#1\ee{\begin{align}#1\end{align}}
\def\nn{\nonumber}
\def\q{\qquad}
\def\f{\frac}
\def\eps{\varepsilon}
\def\teps{\eps}
\def\lb{\big\lbrace}
\def\rb{\big\rbrace}

\def\ip{\lrcorner\,}

\def\tr{\mathrm{tr}}
\def\De{\mathrm{D}}
\def\de{\mathrm{d}}

\def\D{\mathcal{D}}

\def\F{\mathcal{F}}
\def\G{\mathcal{G}}
\def\H{\mathcal{H}}

\def\L{\mathcal{L}}
\def\M{\mathcal{M}}
\def\N{\mathcal{N}}
\def\O{\mathcal{O}}

\def\S{\mathcal{S}}

\def\V{\mathcal{V}}

\numberwithin{equation}{section}

\begin{document}

\title{\Large{\textbf{\sffamily A remarkably simple theory of 3d massive gravity}}}
\author{\sffamily Marc Geiller$^1$ \& Karim Noui$^{2,3}$}
\date{\small{\textit{
$^1$Univ Lyon, ENS de Lyon, Univ Claude Bernard Lyon 1,\\ CNRS, Laboratoire de Physique, UMR 5672, F-69342 Lyon, France\\
$^2$Institut Denis Poisson, Universit\'e de Tours, Universit\'e d'Orl\'eans,\\ CNRS, UMR 7013, 37200 Tours, France\\
$^3$Laboratoire Astroparticule et Cosmologie, Universit\'e Paris Diderot,\\ CNRS, UMR 7164, 75013 Paris, France\\
}}}

\maketitle

\begin{abstract}
We propose and study a new action for three-dimensional massive gravity. This action takes a very simple form when written in terms of connection and triad variables, but the connection can also be integrated out to obtain a triad formulation. The quadratic action for the perturbations around a Minkowski background reproduces the action of self-dual massive gravity, in agreement with the expectation that the theory propagates a massive graviton. We confirm this result at the non-linear level with a Hamiltonian analysis, and show that this new theory does indeed possess a single massive degree of freedom. The action depends on four coupling constants, and we identify the various massive and topological (or massless) limits in the space of parameters. This richness, along with the simplicity of the action, opens a very interesting new window onto massive gravity.
\end{abstract}

\thispagestyle{empty}
\newpage
\setcounter{page}{1}

\hrule
\tableofcontents
\vspace{0.7cm}
\hrule


\newpage

\section{Introduction}

In the past decade, massive gravity has been studied extensively as a possible alternative to general relativity, both at the fundamental and the phenomenological level (see \cite{Hinterbichler:2011tt,Hassan:2011zd,deRham:2014zqa,Schmidt-May:2015vnx} for reviews on massive gravity and bi-metric theories). This interest was triggered in particular by the discovery by de Rham, Gabadaze, and Tolley (dRGT) \cite{deRham:2010kj,deRham:2010ik} of a non-linear theory of massive gravity which propagates the five degrees of freedom\footnote{In this article we will always talk about degrees of freedom in \textit{configuration space}, and not in phase space.} of a massive spin-two particle but does not contain the (in)famous Boulware--Deser ghost \cite{Boulware:1973my}.

At the difference with general relativity, ghost-free massive gravity in four (spacetime) dimensions is not invariant under diffeomorphisms since it requires a ``background'' metric in order to be defined. Diffeomorphism-invariance of dRGT massive gravity can however be restored either by introducing Stueckelberg fields or by considering  the background metric as dynamical. In either case, this restauration is done at the price of including extra dynamical fields in addition to the usual metric.

The situation is radically different in three dimensions, where it is possible to construct non-linear theories of massive gravity which are diffeomorphism-invariant while depending on the degrees of freedom of a single metric only. A first example of such a theory is topologically-massive gravity (TMG), which was introduced in \cite{Deser:1982vy,Deser:1981wh} and whose properties (stability, black hole solutions, holography, etc\dots) have been extensively studied in the literature (see for instance \cite{Solodukhin:2005ah,Li:2008dq,Skenderis:2009nt}). It propagates only one massive graviton (which is possible in three dimensions), breaks parity, and has higher order equations of motion. It also admits a four-dimensional generalization, known as Chern--Simons modified gravity \cite{Alexander:2009tp}, which however breaks Lorentz invariance in addition to parity, and propagates ``only'' three degrees of freedom as a scalar-tensor theory \cite{Crisostomi:2017ugk}. A second example is new massive gravity (NMG), which was introduced in \cite{Bergshoeff:2009hq}. This theory does not break parity, and can be shown to propagate two massive gravitons since it reproduces the Pauli--Fierz action at the linear level. It relies however heavily on the topological nature of three-dimensional gravity, and attempts to generalize it to four dimensions can be shown to lead to the propagation of Ostrogradsky ghosts \cite{Crisostomi:2017ugk}, which makes the resulting theory physically non-viable.

In this article we introduce a new action for massive gravity in three dimensions, given in \eqref{new action} below, which is diffeomorphism- although not Lorentz-invariant, parity-breaking like TMG, but at the difference with other theories of massive gravity does not have higher order equations of motion\footnote{More precisely, the equations of motion are first order, but can also be recast in a second order form.}. This action takes a very simple form and is most naturally written in terms of independent connection and triad variables. More precisely, it is obtained by simply adding to the usual Hilbert--Palatini action a potential term (i.e. with no derivatives) which is invariant under diffeomorphisms but only \textit{global} internal Lorentz transformations. We therefore allow, as the key mechanism, for terms which break the local internal Lorentz symmetry. We argue that there actually exist only two such (potential) terms which can lead to a massive theory of three-dimensional gravity, each coming with a new mass scale. Thus, the new action which we propose contains four coupling constants. These are the Planck mass, the cosmological constant, and the two new mass scales coming with the Lorentz-breaking terms. We show that this action reproduces at the linear level the equations of motion of a model known as self-dual massive gravity \cite{Deser:1984kw,Aragone:1986hm,Deser:1990ay,Bergshoeff:2009hq}, and possesses the single degree of freedom of a (three-dimensional) massive graviton at the full non-linear level. The mass of this graviton depends on the four coupling constants of the new action, which can therefore in a sense be thought of as a describing a four-parameter family of theories of massive gravity.

This article is organized as follows. First, we will present in section \ref{sec2} the new action for three-dimensional massive gravity, and motivate its construction by explaining how (just enough) degrees of freedom can be introduced in three-dimensional gravity by breaking the local internal Lorentz symmetry while retaining diffeomorphism-invariance. In section \ref{sec3} we will then analyse in details some important properties of this new action. We will start by studying the equations of motion and the conditions under which they admit Minkowski spacetime solutions. Next, we will explain how to go from the formulation in terms of independent connection-triad variables to a pure triad formulation where the connection degrees of freedom have been integrated out\footnote{We refrain from calling this a metric formulation since the true dynamical variable will be the triad, and violation of Lorentz invariance will prevent us from rewriting all the terms of the action in terms of $g_{\mu\nu}=e^i_\mu e^j_\nu\eta_{ij}$ (which is obviously a Lorentz-invariant quantity).}. Then, we will study the linearized theory around a Minkowski background, and show precisely how the quadratic action for the perturbations around Minkowski reproduces the action of self-dual massive gravity, in agreement with the expectation that the theory propagates one massive graviton. Finally, we will carry out in section \ref{sec4} the rigorous Hamiltonian analysis of the new action in order to confirm, at the full non-linear level, that it describes the propagation of the single degree of freedom of a three-dimensional massive graviton. We will conclude in section \ref{sec5} with a discussion of interesting open questions and the possible extension of this result to four dimensions. Details about our notations and conventions can be found in appendix \ref{appendixNOT}. Assorted technical comments and results are given in subsequent appendices.

\section{A new action for three-dimensional massive gravity}
\label{sec2}

In this section we present the new action without entering into the technical details, and spend some time discussing the physical motivations behind its construction.

\subsection{Main results}

The context of this work is gravity expressed in the so-called first order formalism, where the dynamical variables are a triad and a gauge connection. These can be seen as $\mathfrak{so}(2,1)$-valued one-forms with respective components $e_\mu^i$ and $\omega_\mu^i$ (see appendix \ref{appendixNOT} for details about our notations). Using the compact notation of differential forms, the new action for three-dimensional massive gravity which we set out to study is
\be\label{new action}
S(e,\omega)=m_\text{p}\int e\wedge\de\omega+\f{\lambda_0}{6}e\wedge[e\wedge e]+\f{\lambda_1}{2}\omega\wedge[e\wedge e]+\f{\lambda_2}{2}e\wedge[\omega\wedge\omega]+\f{\lambda_3}{6}\omega\wedge[\omega\wedge\omega],
\ee
where $m_\text{p}$ is the three-dimensional Planck mass. One can think of this action as being the sum of a kinetic term, which contains the only (first order) derivative, and a potential. The numerical factors have been chosen strictly for later convenience, and will turn out to be very natural. For the unfamiliar reader, when making explicit the spacetime and internal Lorentz indices the action becomes \eqref{new action indices}. In addition to the Planck mass (which is in fact irrelevant when we consider the classical theory without coupling to external matter), this theory depends on the four coupling constants $\lambda_n$, with $n\in\{0,1,2,3\}$.

Since $e$ is interpreted as a dimensionless triad, $\omega$ has the dimension of a mass, and $\lambda_n$ has the dimension of a mass to the power $2-n$. In general relativity, which is obtained when setting $\lambda_1=\lambda_3=0$, the standard coupling constants remain. These are the cosmological constant $\lambda_0$, which has the dimension of a squared mass, and $\lambda_2$ which is dimensionless and can be set to $\lambda_2=1$ without loss of generality\footnote{As can be seen in the action \eqref{new action}, it will be interesting for our purposes to keep $\lambda_2$ explicit. This is because, regardless of whether the coupling constants satisfy the ``massive condition'' $\lambda_0\lambda_3\neq\lambda_1\lambda_2$ or the ``topological condition'' $\lambda_0\lambda_3=\lambda_1\lambda_2$, the limiting case $\lambda_2=0$ is viable and non-trivial. Notice that one can also recover general relativity by taking $\lambda_0=\lambda_2=0$, in which case the roles of $e$ and $\omega$ have to be exchanged and, in particular, $\omega/m_\text{p}$ has to be interpreted as the new triad.}. In the new theory, the coupling constants $\lambda_1$ and $\lambda_3$ are generically non-vanishing and bring in two new mass scales. 

We are going to study the most important properties of the theory \eqref{new action} in great details. In particular, we will show that when the condition $\lambda_0\lambda_3=\lambda_1\lambda_2$ holds, this theory has only first class constraints and does not possess any local degrees of freedom. This topological property in itself is already a surprise, and we comment further on the reason for this in appendix \ref{appendixTOP}. More interesting is the case $\lambda_0\lambda_3\neq\lambda_1\lambda_2$, in which, as we will show, this simple action propagates a single degree of freedom, and as such describes a massive graviton. We will study later on linear perturbations around a Minkowski background, and show how the mass of the graviton depends on the coupling constants.

\subsection{Motivations for breaking internal Lorentz invariance}
\label{sec3.1}

We will now take a step back and first give some motivations leading to the action \eqref{new action}. For this, let us start by recalling basic properties of general relativity with a cosmological constant $\lambda_0$ in three dimensions. The action in this case is given by \eqref{new action} with $\lambda_1=\lambda_3=0$, which is the Hilbert--Palatini action
\be\label{action V0}
S_\text{GR}(e,\omega)=m_\text{p}\int e\wedge\de\omega+V_\text{GR}(e,\omega),
\q
V_\text{GR}(e,\omega)=\f{1}{2}e\wedge[\omega\wedge\omega]+\f{\lambda_0}{6}e\wedge[e\wedge e],
\ee
where in addition we have set $\lambda_2=1$ and made explicit the separation between the kinetic term and the potential. The kinetic term reveals that the spatial components $e^i_a$ and $\omega^i_a$ of the connection and the triad are the only dynamical variables and play the role of canonically-conjugated variables (up to a factor of $\teps^{ab}$). The remaining temporal components, i.e. $e^i_0$ and $\omega^i_0$, are Lagrange multipliers which enforce the six primary constraints. These constraints are first class and generate the six-dimensional symmetry algebra of the theory, which is nothing but $\text{Diff}(M)\times\mathfrak{so}(2,1)$, where $M$ is the spacetime manifold. As a result and as is well-known, the theory described by \eqref{action V0}, i.e. three-dimensional gravity, is topological and has no propagating degrees of freedom.

A natural way of modifying this theory such that it propagates degrees of freedom is to break some of its symmetries. The problem however is that in general breaking the symmetries can lead to the propagation of ghost-like degrees of freedom, which is obviously not desirable. We therefore look for a ``gentle'' breaking of the symmetries which does not introduce ghosts. In the well-known dRGT formulation of massive gravity for instance, diffeomorphisms and internal Lorentz invariance are broken by adding to the action \eqref{action V0} a potential of the form
\be\label{dRGT}
V_\text{dRGT}(e)=\alpha_1e\wedge[f\wedge f]+\alpha_2f\wedge[e\wedge e],
 \ee 
where $\alpha_1$  and $\alpha_2$ are coupling constants, and $f^i_\mu$ is an external fixed triad whose presence clearly breaks diffeomorphism-invariance as well as the internal Lorentz symmetry. This theory has been extensively studied (mostly in four dimensions) and can be shown to propagate in three dimensions two massive gravitons. Indeed, since by definition the potential \eqref{dRGT} does not modify the kinetic term of \eqref{action V0}, only the 12 components $e^i_a$ and $\omega^i_a$ are (canonically-conjugated) dynamical variables, whereas $e^i_0$ and $\omega^i_0$ are Lagrange multipliers enforcing six primary constraints which are now obviously second class. Furthermore, the dRGT potential has been designed in such a way that the theory admits two extra second class constraints, which is a highly non-trivial feature. At the end of the day, glossing over the details of this analysis, the theory propagates $(12-(3+3+2))/2=2$ degrees of freedom. By expanding the action around a Minkowski background when $f_\mu^i=\delta_\mu^i$ is itself a flat triad, these can be shown to represent massive gravitons.

Here, we propose to construct a theory of massive gravity by adding to \eqref{action V0} a potential term $V(e,\omega)$ which is invariant under diffeomorphisms and does not contain new fields. This therefore leaves us with the possibility of breaking only the internal local Lorentz symmetry, while keeping global Lorentz invariance. The most general potential satisfying this requirement can  be expanded in powers of $\omega^i_\mu$ as follows:
\be\label{generic V}
V(e,\omega)=|e|\sum_{\alpha,n}\hat{V}_{i_1\dots i_n}^{j_1\dots j_n}(\alpha)\big(\omega_{\mu_1}^{ i_1}\dots\omega_{\mu_n}^{ i_n}\big)\big(\hat{e}^{\mu_1}_{j_1}\dots\hat{e}^{\mu_n}_{j_n}).
\ee
Here the spacetime indices of the connection components are contracted with that of the inverse triad $\hat{e}$, while internal Lorentz indices are contracted by the tensor $\hat{V}(\alpha)$, which is constructed itself from tensor products of $\eps_{ijk}$ and $\eta_{ij}$. For a given $n$, there are therefore many possible $\hat{V}(\alpha)$'s labelled by $\alpha$. The volume factor $|e|$ is simply ensuring that this is a proper density.

A complete analysis of the degrees of freedom which propagate in the theory obtained by using the potential $V(e,\omega)$ in the action \eqref{action V0} is rather involved in general. However, since the potential does not modify the kinetic structure of the theory, the spatial components $e^i_a$ and $\omega^i_a$ are again canonically-conjugated, and the only subtlety comes from dealing with the components $e^i_0$ and $\omega^i_0$. These can indeed appear in an arbitrary $V(e,\omega)$ in a non-linear manner, which implies that they cannot be treated as Lagrange multipliers. In fact, a quick analysis indicates that with a generic potential there will be too many degrees of freedom, meaning that the resulting theory cannot be considered as a candidate for massive gravity. The argument goes as follows. First, introducing canonical momenta for $e^i_0$ and $\omega^i_0$, we have that the non-physical phase space is spanned by the canonical pairs $(e^i_a,\omega^i_a)$, $(e^i_0,p_i)$, and $(\omega^i_0,\pi_i)$, which is a total of 24 variables. Then, one has to impose the $3+3$ primary constraints $p_i\approx0$ and $\pi_i\approx0$. Because of diffeomorphism-invariance, the preservation of these constraints implies in turn the existence of 3 first class constraints generating diffeomorphisms and another $s$ (secondary, or potentially higher order) constraints. If there are no hidden or accidental symmetries, these $9+s$ constraints will separate into 6 first class constraints and $3+s$ second class ones. At the end of the day, there are therefore $(24-(2\times6+3+s))/2=(9-s)/2$ physical degrees of freedom. While it is of course possible that there exists a clever choice of potential which gives $s=7$, and therefore a single degree of freedom, this requires the existence of (at least) tertiary constraints, and the corresponding theory (with a non-linear dependency on $\omega_0$ and $e_0$) is probably much more complicated than the simple action \eqref{new action} which we propose here. Moreover, in the case $s<7$ the theory with \eqref{generic V} can propagate up to 4 degrees of freedom, and it is likely that some of them are ghosts.

One natural way of getting rid of these extra unwanted degrees of freedom is to consider potentials $V(e,\omega)$ which are at most linear in $e^i_0$ and $\omega^i_0$. In fact, a very similar strategy is implemented in dRGT massive gravity, where one considers potentials which are at most linear in the lapse function and the shift vector. In our case, the most general potential is at most cubic in $\omega^i_\mu$ and takes precisely the form
\be\label{general potential}
V(e,\omega)=\f{\lambda_0}{6}e\wedge[e\wedge e]+\f{\lambda_1}{2}\omega\wedge[e\wedge e]+\f{\lambda_2}{2}e\wedge[\omega\wedge\omega]+\f{\lambda_3}{6}\omega\wedge[\omega\wedge\omega]
\ee
introduced in \eqref{new action}. The two new terms (in addition to the ones defining general relativity) clearly break the internal Lorentz symmetry. At the end of the day, with this potential we obtain the new action which we will now study.

It is interesting to notice that, in this new action \eqref{new action}, the triad $e$ and the connection $\omega$ play a very similar and symmetric role. In fact, it can be seen that the new action satisfies the exchange property
\be
S(e,\omega|\lambda_0,\lambda_1,\lambda_2,\lambda_3)=S(\omega,e|\lambda_3,\lambda_2,\lambda_1,\lambda_0)=S(e',\omega'|m_\text{p}^3\lambda_3,m_\text{p}\lambda_2,m_\text{p}^{-1}\lambda_1,m_\text{p}^{-3}\lambda_0),
\ee
where in the last equality we have defined the new triad $e'\equiv\omega/m_\text{p}$ and the new connection $\omega'\equiv m_\text{p}e$. This is a bit reminiscent of bi-metric theories, such as the ones studied in \cite{Hinterbichler:2012cn,Bergshoeff:2013xma,Alexandrov:2014oda}, although in the present case we are not doubling the number of dynamical variables. The precise link between the action \eqref{new action} and the zwei-Dreibein model of gravity studied in \cite{Alexandrov:2014oda} is given in appendix \ref{appendixBI}.

Finally, a comment about our choice of kinetic term in \eqref{new action} is in order. Indeed, it is possible in principle to add to this action the first order kinetic terms $e\wedge\de e$ and $\omega\wedge\de\omega$, which would have the effect of changing the canonical variables. Here we do not consider this more general possibility in order to keep the symplectic structure of three-dimensional gravity, and so that our massive modification is as minimalistic as possible and consists only in adding the two terms in $\lambda_1$ and $\lambda_3$. Moreover, this choice is justified in appendix \ref{appendixKIN} with a calculation showing that the two extra possible kinetic terms, if initially introduced, can actually be eliminated (under fairly general and reasonable conditions) with a simple change of variables.

\section{Properties of the new action}
\label{sec3}

In this section we study in details some important properties of the new action. First, we compute the equations of motion and give the conditions under which they admit Minkowski spacetime solutions. Then, we explain how to go from a formulation in terms of independent connection and triad variables to a pure triad formulation. Finally, we study the linear analysis around a Minkowski background, and show how the quadratic action for the perturbations reproduces the action of self-dual massive gravity, in agreement with the expectation that the theory propagates one massive graviton.

\subsection{Minkowski solutions}
\label{sec3.1}

Let us start with the first order equations of motion obtained by varying the action \eqref{new action} with respect to $e$ and $\omega$. They are given respectively by
\begin{subequations}\label{EOMs}
\be
\de\omega+\f{\lambda_0}{2}[e\wedge e]+\lambda_1[\omega\wedge e]+\f{\lambda_2}{2}[\omega\wedge\omega]&=0,\\
\de e+\f{\lambda_1}{2}[e\wedge e]+\lambda_2[e\wedge\omega]+\f{\lambda_3}{2}[\omega\wedge\omega]&=0.\label{eom omega}
\ee
\end{subequations}
This shows once again the symmetric role played by the two variables.

We are going to search for flat Minkowski spacetime solutions to these equations of motion. For this, we choose the diagonal Minkowski triad $e^i_\mu=\delta^i_\mu$ and the non-vanishing connection $\omega^i_\mu=\bar{\omega}\delta^i_\mu$, with $\bar{\omega}$ a real constant\footnote{Please note that this variable is different from the one used in appendix \ref{appendixTOP}.} which does not depend on spacetime. Note that this ansatz breaks the symmetry between the role of $e$ and $\omega$. By plugging this in the equations of motion, we find that they reduce to the following two conditions:
\be\label{Mink conditions}
\lambda_0+2\lambda_1\bar{\omega}+\lambda_2\bar{\omega}^2=0,
\q
\lambda_1+2\lambda_2\bar{\omega}+\lambda_3\bar{\omega}^2=0.
\ee
We are now going to classify the solutions to these two equations according to whether $\lambda_2$ and $\lambda_3$ vanish or not. The most generic case corresponds to the situation where $\lambda_2$ and $\lambda_3$ are both non-vanishing. If, in addition, we require the condition $\lambda_1\lambda_3-\lambda_2^2\neq0$, then there is a Minkowski solution only if
\be\label{generic Mink}
\lambda_1\lambda_3-\lambda_2^2\neq0,
\q
4(\lambda_0\lambda_2-\lambda_1^2)(\lambda_1\lambda_3-\lambda_2^2)=(\lambda_0\lambda_3-\lambda_1\lambda_2)^2,
\q
\bar{\omega}=\f{\lambda_1\lambda_2-\lambda_0\lambda_3}{2(\lambda_1\lambda_3-\lambda_2^2)}.
\ee
In the special case where $\lambda_1\lambda_3-\lambda_2^2=0$, one can see immediately that the conditions to have a Minkowski solution 
imply $\lambda_0\lambda_3=\lambda_1\lambda_2$. This particular case will therefore not be so interesting for our analysis, since it corresponds to the topological condition (see appendix \ref{appendixTOP}) in which the theory has no degrees of freedom. For the sake of completeness we can still give the conditions for Minkowski solutions to exist, and these are
\be
\lambda_1\lambda_3-\lambda_2^2=0,
\q
\lambda_0=\f{\lambda_1^2}{\lambda_2}, 
\q
\lambda_3=\f{\lambda_2^2}{\lambda_1},
\q
\bar{\omega}=-\f{\lambda_1}{\lambda_2},
\ee
where we have assumed that $\lambda_1$ is not vanishing neither. Finally, if $\lambda_2=\lambda_3=0$ there is no Minkowski solution (except if $\lambda_0=\lambda_1=0$, in which case the theory becomes trivial).

In order to illustrate and simplify the generic conditions \eqref{generic Mink}, we can consider the four following special cases where only one of the coupling constants $\lambda_n$ vanishes (i.e. $\lambda_2$ and $\lambda_3$ are not simultaneously vanishing):
\begin{itemize}
\item If $\lambda_0=0$, there is a Minkowski solution only if
\be
3\lambda_2^2-4\lambda_1\lambda_3=0,
\q
\bar{\omega}=-2\f{\lambda_1}{\lambda_2}.
\ee
\item If $\lambda_1=0$, there is a Minkowski solution only if
\be
\lambda_0\lambda_3^2+4\lambda_2^3=0,
\q
\bar{\omega}=-2\f{\lambda_2}{\lambda_3}.
\ee
\item If $\lambda_2=0$, there is a Minkowski solution only if
\end{itemize}
\be
4\lambda_1^3+\lambda_0^2\lambda_3=0,
\q
\bar{\omega}=-\f{\lambda_0}{2\lambda_1}.
\ee
\begin{itemize}
\item If $\lambda_3=0$, there is a Minkowski solution only if
\end{itemize}
\be\label{cond l30}
4\lambda_0\lambda_2-3\lambda_1^2=0,
\q
\bar{\omega}=-\f{\lambda_1}{2\lambda_2}.
\ee
General relativity ($\lambda_2=1$ and $\lambda_1=\lambda_3=0$) belongs to this last case, and we recover the condition that a Minkowski solution exists only if there is no cosmological constant, i.e. if $\lambda_0=0$.

Note that this analysis only gives us conditions on the coupling constants for Minkowski spacetime solutions to exist, but does not constraint the theory outside of this sector. For example, when $\lambda_2=\lambda_3=0$ there is no Minkowski solution, but the action \eqref{new action} still defines a non-trivial topological theory (since we necessarily have $\lambda_0\lambda_3=\lambda_1\lambda_2=0$ in this case). Moreover, as we will show in section \ref{sec4}, the theory always has a single degree of freedom when $\lambda_0\lambda_3\neq\lambda_1\lambda_2$, even if for some values of the parameters there may not exist a Minkowski solution.

It would therefore be interesting to extend this analysis and to find the conditions for the theory to admit other physically-relevant solutions, such as de Sitter, anti-de Sitter, and black hole spacetimes. It is however interesting to note at this point that the search for solutions is more subtle than in general relativity owing to the fact that the theory \eqref{new action} is not Lorentz-invariant. This means that, given a triad $e$ which is a solution of \eqref{EOMs} (and the corresponding connection) and which gives a metric $g_{\mu\nu}=e^i_\mu e^j_\nu\eta_{ij}$, a triad $\tilde{e}$ obtained from a Lorentz transformation of $e$ will not necessarily be a solution anymore, although it will of course represent the same metric $g_{\mu\nu}$. The same subtlety appears in other Lorentz-violating theories formulated in terms of triads (or tetrads), such as $f(T)$ teleparallel theories of gravity \cite{Li:2010cg,Tamanini:2012hg}. We will come back to this in future work.

\subsection{Triad formulation}
\label{sec3.2}

In general relativity (i.e. when $\lambda_1=\lambda_3=0$ and $\lambda_2=1$), going from the connection-triad formulation to the metric formulation relies on expressing $\omega$ as a function of $e$ by solving the torsion-free condition \eqref{eom omega}. When $e$ is invertible, this equation has a unique solution given by the Levi--Civita connection, and plugging this solution into the connection-triad Hilbert--Palatini action then leads to the Einstein--Hilbert action. This latter therefore depends only on the triad, or equivalently on the metric through $g_{\mu\nu}=e^i_\mu e^j_\nu\eta_{ij}$.

We can try to follow this method to derive a pure triad action for the modified theory of gravity \eqref{new action}, which requires solving \eqref{eom omega} for $\omega$ for arbitrary values of the coupling constants $\lambda_n$. However, when $\lambda_3\neq0$ this equation is of order two in $\omega$, which makes its resolution only implicit. Because of this difficulty, we are going to first derive the triad action in the case\footnote{The linearization of the theory and the Hamiltonian analysis will of course be performed in the case $\lambda_3\neq0$.} $\lambda_3=0$, and present in appendix \ref{appendix6} the first correction to this result in a perturbative expansion for a small $\lambda_3$.

It turns out that the manipulations involved in this derivation are much more convenient when changing variables and trading $e$ and $\omega$ for the two $3\times3$ matrices
\be\label{Omega and M}
\Omega^{ij}\equiv\teps^{\mu\nu\rho}e^i_\mu\partial_\nu e^j_\rho,
\q
M^{ij}\equiv\omega^i_\mu\hat{e}^{\mu j},
\ee
where $\hat{e}$ is the inverse of $e$ in the sense that $e^i_\mu\hat{e}^\mu_j=\delta^i_j$ and $e^i_\mu\hat{e}^\nu_i=\delta^\nu_\mu$. When working with these matrices the spacetime indices are traded for internal Lie algebra indices only, and we show in appendix \ref{appendixMAT} that the action \eqref{new action} can be written as
\be\label{matrix action}
S(e,\omega)=m_\text{p}\int\de^3x\,\Big\{\tr(\Omega M)+V(M)\Big\},
\ee
where the potential $V(M)$ is given by
\be
V(M)=-|e|\left(\lambda_0+\lambda_1\tr(M)+\f{\lambda_2}{2}\big[\tr^2(M)-\tr(M^2)\big]+\lambda_3\det(M)\right),
\ee
and where $\det(M)$ is the determinant of the matrix ${M^i}_j$ (with indices up and down). We now want to study the equations of motion, and in the case $\lambda_3=0$ go from the connection-triad to the triad formulation by writing $M$ as a function of $\Omega$. Taking $\lambda_3=0$, we have that the equations of motion \eqref{eom omega} written in terms of $\Omega$ and $M$ take the form
\be\label{eom omega matrices}
\Omega-|e|\Big(\lambda_1\eta+\lambda_2\big[\tr(M)\eta-M\big]\Big)=0.
\ee
Taking the trace of this equation then leads to
\be
\tr(M)=\f{1}{2\lambda_2|e|}\tr(\Omega)-\f{3\lambda_1}{2\lambda_2},
\ee
which when plugged back into \eqref{eom omega matrices} gives the solution
\be\label{solution for M}
M=\f{1}{2\lambda_2|e|}\big[\tr(\Omega)\eta-2\Omega\big]-\f{\lambda_1}{2\lambda_2}\eta.
\ee
This equation is essentially the solution of the equations of motion \eqref{eom omega} (still in the case $\lambda_3=0$) which gives $\omega$ in terms of $e$, and explicit expressions for the connection are given in appendix \ref{appendixSOL}. 
Now, we can also multiply \eqref{eom omega matrices} by $M$ before taking the trace to find
\be
\lambda_2|e|\big[\tr^2(M)-\tr(M^2)\big]=\tr(\Omega M)-\lambda_1|e|\tr(M),
\ee
and multiply \eqref{solution for M} by $\Omega$ before taking the trace to find
\be
\tr(\Omega M)=\f{1}{2\lambda_2|e|}\big[\tr^2(\Omega)-2\tr(\Omega^2)\big]-\f{\lambda_1}{2\lambda_2}\tr(\Omega).
\ee
Inserting these expressions into the original action \eqref{matrix action} finally leads to the matrix form of the triad action, which is
\be\label{second order l3=0}
S_0(e)=\f{m_\text{p}}{2\lambda_2}\int\de^3x\left\{\f{1}{2|e|}\big[\tr^2(\Omega)-2\tr(\Omega^2)\big]+\f{1}{2}(3\lambda_1^2-4\lambda_0\lambda_2)|e|-\lambda_1\tr(\Omega)\right\}.
\ee
At this stage the action is still written in terms of the triad, and it is natural to ask whether it is possible to write it in terms of the metric $g_{\mu\nu}=e^i_\mu e^j_\nu\eta_{ij}$.

To go from the triad formulation to an expression involving the metric, we can first switch between the fundamental and the adjoint representation of the Lie algebra by introducing the notation $\omega^{ij}_\mu\equiv-{\eps^{ij}}_k\omega^k_\mu$ and writing
\be
R^{ij}_{\mu\nu}=\partial_\mu\omega^{ij}_\nu-\partial_\nu\omega^{ij}_\mu+\omega^i_{\mu k}\omega^{kj}_\nu-\omega^i_{\nu k}\omega^{kj}_\mu=-{\eps^{ij}}_k\big(\partial_\mu\omega^k_\nu-\partial_\nu\omega^k_\mu+{\eps^k}_{mn}\omega^m_{\mu}\omega^n_\nu\big)=-{\eps^{ij}}_kF^k_{\mu\nu},
\ee
where eventually we will take $\omega^{ij}$ to be the torsion-free connection $\Gamma^{ij}(e)$ given below \eqref{torsion-free connection}, so that this $R_{\mu\nu}$ becomes the Ricci tensor. Use the identity $\teps^{\mu\nu\rho}\eps_{ijk}e^k_\rho=-e(\hat{e}^\mu_i\hat{e}^\nu_j-\hat{e}^\mu_j\hat{e}^\nu_i)$ to write
\be
\eta_{ij}\teps^{\mu\nu\rho}e^i_\mu F^j_{\nu\rho}=\f{1}{2}\teps^{\alpha\mu\nu}\eps_{ijk}e^i_\alpha R^{jk}_{\mu\nu}=-e\hat{e}^\mu_i\hat{e}^\nu_jR^{ij}_{\mu\nu}=-\sqrt{g}R,
\ee
we then get that
\be\label{metric second order l3=0}
S_0(e)=-\f{m_\text{p}}{2\lambda_2}\int\de^3x\left\{\sqrt{|g|}(R-2\Lambda)+\lambda_1\eps^{\mu\nu\rho}e^i_\mu\partial_\nu e_{\rho i}\right\}.
\ee
When $\lambda_1=0$ we therefore recover the Einstein--Hilbert action with a cosmological constant equal to $\Lambda\equiv(3\lambda_1^2-4\lambda_0\lambda_2)/4$. This is indeed the combination of coupling constants which has to vanish in order for Minkowski spacetime to be a solution, in agreement with the case \eqref{cond l30} discussed above. Finally, one can see that the parity-breaking term coming with $\lambda_1$ cannot be rewritten in terms of the metric because of the pattern of index contraction.

\subsection[Linearization for $\lambda_3=0$]{Linearization for $\boldsymbol{\lambda_3=0}$}
\label{sec3.3}

We are now going to study the linearization of the theory around a Minkowski background, which will exhibit and make manifest the presence of the massive graviton. With the Hamiltonian analysis we will then confirm the presence of this single degree of freedom at the full non-linear level.

Before presenting the general result, let us first focus on the simpler case $\lambda_3=0$ as in the previous subsection. In this case, we have obtained in \eqref{second order l3=0} the triad action, which we can take as our starting point. We consider perturbations around a Minkowski background by writing
\be
e^i_\mu=\delta^i_\mu+f^i_\mu,
\ee
and expand the action \eqref{second order l3=0} to second order in $f$. Plugging this expression for the linearized triad in the definition \eqref{Omega and M} of $\Omega$ leads to
\begin{subequations}\label{linear Omega}
\be
\Omega^{ij}&=\teps^{\mu\nu\rho}(\delta^i_\mu+f^i_\mu)\partial_\nu f^j_\rho,\\
\tr(\Omega)&=\teps^{\mu\nu\rho}(\partial_\rho f_{\mu\nu}+{f_\mu}^\sigma\partial_\nu f_{\rho\sigma}),\\
\tr^2(\Omega)&=\teps^{\mu\nu\rho}\teps^{\alpha\beta\sigma}(\partial_\rho f_{\mu\nu})(\partial_\sigma f_{\alpha\beta})+\O(f^3),\\
\tr(\Omega^2)&=\teps^{\mu\nu\rho}\teps^{\alpha\beta\sigma}(\partial_\nu f_{\rho\alpha})(\partial_\beta f_{\sigma \mu})+\O(f^3),
\ee
\end{subequations}
where we have introduced the notation $f_{\mu\nu}\equiv f^i_\mu\delta_{\nu i}$. Notice that this $f_{\mu\nu}$ is therefore \textit{not} symmetric. Using \eqref{linear Omega} in the action \eqref{second order l3=0} then leads to the following quadratic action for the perturbations:
\be\label{quadratic action}
S_0(f)
&=\f{m_\text{p}}{2\lambda_2}\int\de^3x\left\{\teps^{\mu\nu\rho}\teps^{\alpha\beta\sigma}\left(\f{1}{2}(\partial_\rho f_{\mu\nu})(\partial_\sigma f_{\alpha\beta})-(\partial_\nu f_{\rho\alpha})(\partial_\beta f_{\sigma \mu})\right)-\lambda_1\teps^{\mu\nu\rho}{f_\mu}^\sigma\partial_\nu f_{\rho\sigma}\right\}\nn\\
&=\f{m_\text{p}}{2\lambda_2}\int\de^3x\left\{\left(\f{1}{2}\teps^{\mu\nu\rho}\teps^{\alpha\beta\sigma}-\teps^{\mu\beta\rho}\teps^{\nu\sigma\alpha}\right)(\partial_\rho f_{\mu\nu})(\partial_\sigma f_{\alpha\beta})-\lambda_1\teps^{\mu\nu\rho}{f_\mu}^\sigma\partial_\nu f_{\rho\sigma}\right\},
\ee
where the second rewriting will be useful when deriving the equations of motion. At the linear level, our new theory \eqref{new action} therefore reproduces the action of self-dual massive gravity \cite{Deser:1984kw,Aragone:1986hm,Deser:1990ay,Bergshoeff:2009hq}, exactly as dRGT reproduces the Pauli--Fierz theory in four dimensions.

Let us make two comments before studying the equations of motion of this action for the perturbations. First, one can see that this action is invariant under the linearized diffeomorphisms defined by the transformation law $\delta_\xi f_{\mu\nu}=\partial_\mu \xi_\nu$ for any one-form $\xi_\mu$. Second, decomposing the perturbations $f_{\mu\nu}$ into a symmetric part $h_{\mu\nu}=h_{\nu\mu}$ and an anti-symmetric part described by a vector $A^\rho$ as
\be\label{decomposition of f}
f_{\mu\nu}=h_{\mu\nu}+\teps_{\mu\nu\rho}A^\rho,
\ee
one gets in the case $\lambda_1=0$ corresponding to general relativity that
\be
S_\text{GR}(h)=-\f{m_\text{p}}{2\lambda_2}\int\de^3x\,\teps^{\mu\nu\rho}\teps^{\alpha\beta\sigma}(\partial_\nu h_{\rho\alpha})(\partial_\beta h_{\sigma \mu}).
\ee
This is the usual linearized action for the metric perturbations $h_{\mu\nu}$. Note that in this calculation the relative coefficient of $-2$ between the first two terms in \eqref{quadratic action} is crucial in order to get the expected result. Indeed, any other coefficient would have left in the action a term of the form $(\partial_\mu A^\mu)^2$, which would be responsible for the propagation of an extra ghost-like degree of freedom (for the longitudinal mode of the vector introduced as $A_\mu=\partial_\mu\phi$).

We are now going to study the equations of motion for the perturbations obtained from \eqref{quadratic action}, which is known to reproduce the dynamics of a massive graviton. Here, we would like to manipulate the equations of motion to arrive at equations which explicitly suggest that a massive graviton is propagating. For this, we start by differentiating this action with respect to $f_{\mu\nu}$, which leads to the equations of motion
\be
\left(\f{1}{2}\teps^{\mu\nu\rho}\teps^{\sigma\alpha\beta}-\teps^{\mu\beta\rho}\teps^{\nu\sigma\alpha}\right)\partial_\rho\partial_\sigma f_{\alpha\beta}+\lambda_1\teps^{\mu\rho\sigma}\partial_\rho{f_\sigma}^\nu=0.
\ee
Expanding the two products of Levi--Civita symbols, one gathers symmetric combinations of terms, which can be written in terms of $h_{\mu\nu}$, and these equations of motion become
\be\label{eom for h}
\Box h_{\mu\nu}+\partial_\mu\partial_\nu h-\Box h\eta_{\mu\nu}+(\partial^\rho\partial^\sigma h_{\rho\sigma})\eta_{\mu\nu}-\partial^\rho(\partial_\mu h_{\rho\nu}+\partial_\nu h_{\rho\mu})+\lambda_1\teps_{\mu\rho\sigma}\partial^\rho{f^\sigma}_\nu=0.
\ee
We can now manipulate this expression in several ways to obtain useful relations. First, multiplying by $\eta^{\mu\nu}$ and $\teps^{\mu\nu\alpha}$ leads respectively to
\begin{subequations}
\be
\Box h-\partial^\rho\partial^\sigma h_{\rho\sigma}-\lambda_1\teps_{\mu\rho\sigma}\partial^\rho f^{\sigma\mu}=0,
\q
\partial^\alpha f-\partial^\nu{f^\alpha}_\nu=0,
\ee
\end{subequations}
where $h\equiv {h^\mu}_\mu$. Then, acting on the second relation with $\partial_\alpha$ and making use of the decomposition \eqref{decomposition of f} leads to $\Box h-\partial^\rho\partial^\sigma h_{\rho\sigma}=0$, so that the first relation reduces to $\teps_{\mu\rho\sigma}\partial^\rho f^{\sigma\mu}=0$. Using once again the decomposition \eqref{decomposition of f} in this finally gives
\be\label{gauge for A}
\partial_\mu A^\mu=0,
\ee
which we can supplement by the gauge condition
\be\label{gauge for f}
\partial^\mu f_{\mu\nu}=\partial^\mu h_{\mu\nu}+\teps_{\mu\nu\rho}\partial^\mu A^\rho=0.
\ee
Using these relations we can now simplify the equations of motion \eqref{eom for h} to put them in the form
\be
\Box h_{\mu\nu}+\partial_\mu\partial_\nu h-\partial^\rho(\partial_\mu h_{\rho\nu}+\partial_\nu h_{\rho\mu})+\lambda_1\teps_{\mu\rho\sigma}\partial^\rho{h^\sigma}_\nu+\lambda_1\partial_\nu A_\mu=0.
\ee
Multiplying this equation by $\teps^{\mu\nu\alpha}$ and using the gauge condition \eqref{gauge for f} now leads to
\be\label{relation for h}
\partial_\alpha h=2\partial^\rho h_{\rho\alpha},
\ee
which can be used in the equations of motion to simply them further and obtain
\be
\Box h_{\mu\nu}+\lambda_1\teps_{\mu\rho\sigma}\partial^\rho{h^\sigma}_\nu+\lambda_1\partial_\nu A_\mu=0.
\ee
Multiplying this by $\eta^{\mu\nu}$ and using \eqref{gauge for A} now gives $\Box h=0$, while acting with $\partial^\nu$ and using \eqref{relation for h} leads to $\Box A_\mu=0$. With this, we can then act on the equations of motion with $\Box$ to obtain
\be
\Box^2h_{\mu\nu}+\lambda_1\teps_{\mu\rho\sigma}\partial^\rho\Box{h^\sigma}_\nu=\Box^2h_{\mu\nu}-\lambda_1^2\teps_{\mu\rho\sigma}\partial^\rho(\teps^{\sigma\alpha\beta}\partial_\alpha h_{\beta\nu}+\partial_\nu A^\sigma)=0.
\ee
Finally, expanding the Levi--Civita symbol and using \eqref{gauge for f} together with \eqref{relation for h} leads to the result
\be
\Box\big(\Box-\lambda_1^2\big)h_{\mu\nu}=0.
\ee
This propagation equation, which is slightly unusual since it has an extra d'Alembertian operator, strongly suggests (but does not prove) that the dynamical degree of freedom is a massive graviton of mass $\lambda_1$. More precisely, this equation does in fact tell us that the theory contains a massless and/or a massive excitation. However, massless gravitons do not propagate in three dimensions. Therefore, since the Hamiltonian analysis reveals that the theory has a single propagating degree of freedom (at the full non-linear level), this necessarily means that it describes a massive graviton, in agreement with the analysis of \cite{Deser:1982vy,Deser:1981wh} or more recently of \cite{Bergshoeff:2009hq}.

\subsection[Linearization for $\lambda_3\neq0$]{Linearization for $\boldsymbol{\lambda_3\neq0}$}
\label{sec3.4}

We can now generalize the result of the previous subsection to the case $\lambda_3\neq0$. Since we do not have the expression for the pure triad action in this case (appart from the perturbative result of appendix \ref{appendix6}), we are going to linearize the connection-triad action \eqref{new action} instead. In order to linearize this action around a Minkowski background, we expand the triad and the connection as
\be
e^i_\mu=\delta^i_\mu+f^i_\mu,
\q
\omega^i_\mu=\bar{\omega}\delta^i_\mu+q^i_\mu.
\ee
First, plugging this in the action \eqref{new action} leads to
\be
S(f,q)&=m_\text{p}\int\de^3x\left\{\eps^{\mu\nu\rho}f_{\mu\sigma}\partial_\nu{q_\rho}^\sigma+\f{1}{2}(\lambda_0+\lambda_1\bar{\omega})(f_{\mu\nu}f^{\nu\mu}-f^2)+\f{1}{2}(\lambda_2+\lambda_3\bar{\omega})(q_{\mu\nu}q^{\nu\mu}-q^2)\right.\nn\\
&\phantom{\ =m_\text{p}\int\de^3x\left\{\eps^{\mu\nu\rho}f_{\mu\sigma}\partial_\nu{q_\rho}^\sigma\right.}\left.+(\lambda_1+\lambda_2\bar{\omega})(f_{\mu\nu}q^{\nu\mu}-fq)\vphantom{\f{1}{2}}\right\},
\ee
where we have again used the notation $f_{\mu\nu}\equiv f^i_\mu\delta_{\nu i}$. Then, using the conditions \eqref{Mink conditions} for the Minkowski solution enables us to rewrite this action in the form
\be
S(f,q)
&=m_\text{p}\int\de^3x\left\{\eps^{\mu\nu\rho}f_{\mu\sigma}\partial_\nu{q_\rho}^\sigma+\f{1}{2}(\lambda_2+\lambda_3\bar{\omega})\Big((q_{\mu\nu}-\bar{\omega}f_{\mu\nu})(q^{\mu\nu}-\bar{\omega}f^{\mu\nu})-(q-\bar{\omega}f)^2\Big)\right\}\nn\\
&=m_\text{p}\int\de^3x\left\{\eps^{\mu\nu\rho}(p_{\mu\sigma}+\bar{\omega}f_{\mu\sigma})\partial_\nu{f_\rho}^\sigma+\f{1}{2}(\lambda_2+\lambda_3\bar{\omega})(p_{\mu\nu}p^{\mu\nu}-p^2)\right\},\label{linearized first order action}
\ee
where for the second equality we have introduced the new variable
\be
p_{\mu\nu}\equiv q_{\mu\nu}-\bar{\omega}f_{\mu\nu}.
\ee
This is the linearized connection-triad action for arbitrary values of the coupling constants. While it was not possible in section \ref{sec3.2} to obtain the triad action for the non-linear theory with $\lambda_3\neq0$, at the linearized level this calculation is however possible.

To obtain the triad action for the perturbations, we have to proceed like in the non-linear case and solve for half of the equations of motion. The equations of motion obtained by differentiating with respect to $p_{\mu\sigma}$ are
\be
\teps^{\mu\nu\rho}\partial_\nu{f_\rho}^\sigma+(\lambda_2+\lambda_3\bar{\omega})(p^{\sigma\mu}-p\eta^{\sigma\mu})=0.
\ee
This can be solved to find
\be
p^{\sigma\mu}=\f{1}{2(\lambda_2+\lambda_3\bar{\omega})}(\eta^{\sigma\mu}\eps^{\alpha\beta\rho}\partial_\rho f_{\alpha\beta}-2\eps^{\mu\nu\rho}\partial_\nu{f_\rho}^\sigma),
\ee
where we further assume that $\lambda_2+\lambda_3\bar{\omega}\neq0$. This can finally be inserted back into the linearized connection-triad action \eqref{linearized first order action} (after first contracting the equations of motion with $p_{\sigma\mu}$ to simplify the action) to find the triad action
\be\label{general quadratic action}
S(f)=\f{m_\text{p}}{2(\lambda_2 + \lambda_3 \bar{\omega})}\int\de^3x\left\{\left(\f{1}{2}\teps^{\mu\nu\rho}\teps^{\alpha\beta\sigma}-\teps^{\mu\beta\rho}\teps^{\nu\sigma\alpha}\right)(\partial_\rho f_{\mu\nu})(\partial_\sigma f_{\alpha\beta})-m_\text{g}\teps^{\mu\nu\rho}{f_\mu}^\sigma\partial_\nu f_{\rho\sigma}\right\},
\ee
where we have introduced the new mass
\be\label{graviton mass}
m_\text{g}\equiv-2\bar{\omega}(\lambda_2+\lambda_3\bar{\omega}),
\ee
and where the dependency of $\bar{\omega}$ on the coupling constants is determined by the different cases discussed in section \ref{sec3.1}.
It is important to emphasize once again that the expression \eqref{general quadratic action} for the quadratic action is valid only if 
$\lambda_2+\lambda_3\bar{\omega}\neq0$. If this condition is not satisfied the quadratic action \eqref{linearized first order action} trivializes, which is a sign of a strong coupling problem.

The remarkable result \eqref{general quadratic action}, which extends naturally that of the previous subsection, shows that for any values of the coupling constants $\lambda_n$ compatible with the massive condition $\lambda_0\lambda_3\neq\lambda_1\lambda_2$, the (no-strong coupling) condition
$\lambda_2+\lambda_3\bar{\omega}\neq 0$, and the criteria of section \ref{sec3.1}, we obtain the same quadratic action \eqref{general quadratic action} for the perturbations. This therefore achieves the proof that at the linear level the new action \eqref{new action} describes a massive graviton of mass $m_\text{g}$ determined by the coupling constants.

Let us conclude by looking at specific cases for the mass of the graviton by combining \eqref{graviton mass} with the results of section \ref{sec3.1}. First of all, one can see that the graviton is massless, i.e. that $m_\text{g}=0$, only when $\overline{\omega}=0$. This in turn implies that $\lambda_0=\lambda_1=0$, which corresponds to the topological case in which the theory has no propagating degrees of freedom. Conversely, if the topological condition $\lambda_0\lambda_3=\lambda_1\lambda_2$ is satisfied then $\bar{\omega}=0$ and the graviton is massless (assuming that $\lambda_2+\lambda_3\bar{\omega}\neq0$), as can be seen from the expression \eqref{graviton mass} and the results of section \ref{sec3.1}. As a conclusion, we have as expected an equivalence between the topological sector and the masslessness of the graviton (still bearing in mind that we actually have a family of topological theories since the topological condition can be satisfied in many different ways).

In the generic massive case, when the condition $\lambda_1\lambda_3-\lambda_2^2\neq0$ is satisfied (which is required in order to have a Minkowski solution), the mass is given by $m_\text{g}$ with $\bar{\omega}$ as in \eqref{generic Mink}. Explicitly this is
\be\label{generic mass}
m_\text{g}=\f{2\lambda_1^2\lambda_3-\lambda_1\lambda_2^2-\lambda_0\lambda_2\lambda_3}{\lambda_1\lambda_3-\lambda_2^2}.
\ee
Then we can look at the particular cases studied in section \ref{sec3.1}, where only one of the coupling constants is vanishing, and find that the corresponding masses are given by
\be
m_\text{g}^{(\lambda_0=0)}=-2\lambda_1,
\q
m_\text{g}^{(\lambda_1=0)}=-4\f{\lambda_2^2}{\lambda_3},
\q
m_\text{g}^{(\lambda_2=0)}=2\lambda_1,
\q
m_\text{g}^{(\lambda_3=0)}=\lambda_1.
\ee
Beyond the simple exercise in numerology, these expressions are interesting as a consistency check and as a way to illustrate the subtleties which can arise when sending some of the coupling constants of \eqref{new action} to zero. For example, the limit of $m_\text{g}^{(\lambda_0=0)}=-2\lambda_1$ when $\lambda_2\rightarrow0$ gives $-2\lambda_1$, while starting from a (topological) theory with $\lambda_0=\lambda_2=0$ from the onset leads to a vanishing mass. This indicates that one has to be careful when studying the topological limit of a massive theory (and even more so in the present case since we have a four-parameter family of theories). Many other subtle example can be worked out.

\section{Hamiltonian analysis}
\label{sec4}

We are now going to proceed to the Hamiltonian analysis of the new action \eqref{new action}, which will reveal the role played by the condition $\lambda_0\lambda_3\neq\lambda_1\lambda_2$, and show that when it is satisfied this theory has a single degree of freedom. We follow the usual Dirac algorithm, and therefore start with the primary constraints before evolving them in time to study the secondary constraints. After having gathered all the constraints, we separate them between first and second class, and then proceed to the counting of the degrees of freedom. As announced, this counting will lead to a single configuration space degree of freedom in the case $\lambda_0\lambda_3\neq\lambda_1\lambda_2$, and to zero degrees of freedom otherwise.

\subsection{Primary constraints}

First, by putting the terms in $\lambda_2$ together with ordinary derivatives in order to define the curvature and torsion two-forms
\be
\widetilde{F}\equiv\de\omega+\f{\lambda_2}{2}[\omega\wedge\omega],
\q
\widetilde{\De}e\equiv\de e+\lambda_2[\omega\wedge e],
\ee
one can easily see that the variation of the action is given by\footnote{We neglect possible boundaries and terms obtained from integrations by parts. If boundaries are present, the variational principle and the choice of boundary conditions are the same as in general relativity since we have the standard kinetic term $e\wedge\de\omega$ in our action \eqref{new action}.}
\be\label{action variation}
\delta S=m_\text{p}\int\delta e\wedge\left(\widetilde{F}+\f{1}{2}\big(\lambda_0[e\wedge e]+2\lambda_1[\omega\wedge e]\big)\right)+\delta\omega\wedge\left(\widetilde{\De}e+\f{1}{2}\big(\lambda_1[e\wedge e]+\lambda_3[\omega\wedge\omega]\big)\right).
\ee
This will be useful below in order to compute the action of the symmetries. Notice that in the case $\lambda_2=1$ we have that $\widetilde{F}=F$ and $\widetilde{\De}e=\De e$ correspond to the familiar definitions of curvature and torsion.

Starting from the action \eqref{new action} and making all the indices explicit, one can perform a $2+1$ decomposition of the spacetime indices as $\mu=\{0,a\}$ and write the action in the Hamiltonian form
\be\label{new action indices}
S
&=m_\text{p}\int\de^3x\ \teps^{\mu\nu\rho}\left\{e^i_\mu\partial_\nu\omega_{\rho i}+\eps_{ijk}\left(\f{\lambda_0}{6}e^i_\mu e^j_\nu e^k_\rho+\f{\lambda_1}{2}\omega^i_\mu e^j_\nu e^k_\rho+\f{\lambda_2}{2}e^i_\mu\omega^j_\nu\omega^k_\rho+\f{\lambda_3}{6}\omega^i_\mu\omega^j_\nu\omega^k_\rho\right)\right\}\nn\\
&=m_\text{p}\int\de^3x\,\teps^{ab}\left\{\partial_0\omega^i_ae_{bi}+e^i_0\left(\f{1}{2}\widetilde{F}_{abi}+\f{1}{2}\eps_{ijk}\big(\lambda_0e^j_ae^k_b+2\lambda_1\omega^j_ae^k_b\big)\right)\right.\nn\\
&\phantom{\ =m_\text{p}\int\de^3x\,\teps^{ab}\left\{\partial_0\omega^i_ae_{bi}\right.}\left.+\omega^i_0\left(\widetilde{\De}_ae_{bi}+\f{1}{2}\eps_{ijk}\big(\lambda_1e^j_ae^k_b+\lambda_3\omega^j_a\omega^k_b\big)\right)\right\}\nn\\
&=m_\text{p}\int\de t\int_\Sigma\partial_0\omega\wedge e+e_0\left(\widetilde{F}+\f{1}{2}\big(\lambda_0[e\wedge e]+2\lambda_1[\omega\wedge e]\big)\right)\nn\\
&\phantom{\ =m_\text{p}\int\de t\int_\Sigma\partial_0\omega\wedge e}+\omega_0\left(\widetilde{\De}e+\f{1}{2}\big(\lambda_1[e\wedge e]+\lambda_3[\omega\wedge\omega]\big)\right),
\ee
where the spatial components of the curvature and the torsion are given by
\be
\widetilde{F}^i_{ab}=\partial_a\omega^i_b-\partial_b\omega^i_a+\lambda_2{\eps^i}_{jk}\omega^j_a\omega^k_b,
\q
\widetilde{\De}_ae^i_b=\partial_ae^i_b+\lambda_2{\eps^i}_{jk}\omega^j_ae^k_b.
\ee
In the last equality, we have simply rewritten the $2+1$ decomposition in terms of differential forms on the two-dimensional spatial manifold $\Sigma$. This compact notation is very useful for the rest of the calculations, and there should be no possible confusion since one can clearly see that the integrand only makes sense as a two-form.

From the Hamiltonian form of the action, one can now read the canonical Poisson brackets between the phase space variables:
\be
\lb e^i_a(x),\omega^j_b(y)\rb=\lb\omega^i_a(x),e^j_b(y)\rb=\eta^{ij}\teps_{ab}\delta^{(2)}(x,y).
\ee
The Hamiltonian itself is given by
\be\label{Hamiltonian constraint}
\H=\F(e_0)+\G(\omega_0),
\ee
and is as usual the sum of the primary constraints enforced by the Lagrange multipliers $e^i_0$ and $\omega^i_0$, which are given in smeared form by
\begin{subequations}
\be
\F(\alpha)&\equiv\int_\Sigma\alpha\left(\widetilde{F}+\f{1}{2}\big(\lambda_0[e\wedge e]+2\lambda_1[\omega\wedge e]\big)\right)\approx0,\\
\G(\alpha)&\equiv\int_\Sigma\alpha\left(\widetilde{\De}e+\f{1}{2}\big(\lambda_1[e\wedge e]+\lambda_3[\omega\wedge\omega]\big)\right)\approx0.
\ee
\end{subequations}
The infinitesimal action of these constraints on the phase space variables is given by the Poisson brackets
\begin{subequations}\label{infinitesimal transformations}
\be
\lb\F(\alpha),e\rb&=\widetilde{\De}\alpha+\lambda_1[e,\alpha],\\
\lb\F(\alpha),\omega\rb&=\lambda_0[e,\alpha]+\lambda_1[\omega,\alpha],\\
\lb\G(\alpha),e\rb&=\lambda_2[e,\alpha]+\lambda_3[\omega,\alpha],\\
\lb\G(\alpha),\omega\rb&=\widetilde{\De}\alpha+\lambda_1[e,\alpha],
\ee
\end{subequations}
where once again all these differential forms should be understood as being pulled-back to the spatial hypersurface $\Sigma$.

With this, one can now see that spatial diffeomorphisms are obtained, up to the primary constraints (which are nothing but the spatial components of the equations of motion), as the action of $\F$ and $\G$ with specific field-dependent smearing functions. More precisely, using the notation $\xi\ip v=\xi^av_a$ for a one-form $v$ and for a vector field $\xi\in\Sigma$, we have the following formulas:
\begin{subequations}
\be
\lb\F(\alpha),e\rb\big|_{\alpha=\xi\ip e}+\lb\G(\alpha),e\rb\big|_{\alpha=\xi\ip\omega}+\xi\ip\left(\widetilde{\De}e+\f{1}{2}\big(\lambda_1[e\wedge e]+\lambda_3[\omega\wedge\omega]\big)\right)&=\L_\xi e,\\
\lb\F(\alpha),\omega\rb\big|_{\alpha=\xi\ip e}+\lb\G(\alpha),\omega\rb\big|_{\alpha=\xi\ip\omega}+\xi\ip\left(\widetilde{F}+\f{1}{2}\big(\lambda_0[e\wedge e]+2\lambda_1[\omega\wedge e]\big)\right)&=\L_\xi\omega,
\ee
\end{subequations}
where $\L_\xi(\,\cdot\,)=\de(\xi\ip\,\cdot\,)+\xi\ip(\de\,\cdot\,)$ is the Lie derivative along the vector field $\xi$. This means that the quantity $\V(\xi)\equiv\F(\xi\ip e)+\G(\xi\ip\omega)$ is the generator of the two spatial diffeomorphisms for $\xi\in\Sigma$. As for the generator of time-like diffeomorphisms, it is given by the Hamiltonian constraint \eqref{Hamiltonian constraint} with the values of the multipliers determined by the Hamiltonian analysis. Since $\V$ and $\H$ are built from the same combination of $\F$ and $\G$ and simply feature different smearing fields, we might as well consider a spacetime vector field $X$ and the three smeared constraints
\be
\D(X)\equiv\F(X\ip e)+\G(X\ip\omega),
\ee
where the smearing is now with $X\ip v=X^\mu v_\mu$. These are the generators of spacetime diffeomorphisms, or in other words the three first class constraints which can be extracted from $\F$ and $\G$, and which we expect to find since the theory is manifestly diffeomorphism-invariant.

\subsection{Secondary constraints}

What is now important is to study the time evolution of the primary constraints $\F$ and $\G$.
For this, it is useful to first compute the three elementary Poisson brackets between the constraints. A lengthy but elementary calculation shows that these are given by
\begin{subequations}
\be
\lb\F(\alpha),\F(\beta)\rb
&=\lambda_0\G([\alpha,\beta])+\lambda_1\F([\alpha,\beta])+\f{1}{2}(\lambda_1\lambda_2-\lambda_0\lambda_3)\int_\Sigma[\alpha,\beta][\omega\wedge\omega],\\
\lb\G(\alpha),\G(\beta)\rb
&=\lambda_2\G([\alpha,\beta])+\lambda_3\F([\alpha,\beta])+\f{1}{2}(\lambda_1\lambda_2-\lambda_0\lambda_3)\int_\Sigma[\alpha,\beta][e\wedge e],\\
\lb\F(\alpha),\G(\beta)\rb
&=\lambda_1\G([\alpha,\beta])+\lambda_2\F([\alpha,\beta])+(\lambda_1\lambda_2-\lambda_0\lambda_3)\int_\Sigma[\alpha,e]\wedge[\omega,\beta].
\ee
\end{subequations}
Remarkably, one can see that all these Poisson brackets are weakly vanishing if $\lambda_0\lambda_3=\lambda_1\lambda_2$. In this case, the $3+3$ primary constraints $\F$ and $\G$ are first class, the Dirac algorithm stops, and there are $(12-2\times(3+3))/2=0$ degrees of freedom. The reason behind this topological property is that when the topological condition on the coupling constants is satisfied there is a hidden local Lorentz symmetry in addition to the manifest diffeomorphism symmetry. Yet another way to see this is to promote the infinitesimal action \eqref{infinitesimal transformations} of the constraints on the phase space variables to an action on all the spacetime components of the variables (i.e. to also act on the multipliers). Then we can plug the action of $\F$ and $\G$ in the infinitesimal variation \eqref{action variation} to find
\begin{subequations}
\be
\delta^\F_\alpha S=\f{1}{2}\int(\lambda_1\lambda_2-\lambda_0\lambda_3)[\alpha,e]\wedge[\omega\wedge\omega],
\q
\delta^\G_\alpha S=\f{1}{2}\int(\lambda_1\lambda_2-\lambda_0\lambda_3)[\alpha,\omega]\wedge[e\wedge e].
\ee
\end{subequations}
This again shows that when the topological condition on the coupling constants is satisfied, the action is invariant under the action of $3+3$ symmetries (although it should be clear that neither $\F$ nor $\G$ act like infinitesimal Lorentz transformations), which kills all the degrees of freedom and results in a topological theory. This is explained in more details in appendix \ref{appendixTOP}, where we show that in this topological case there is a change of variables which maps the action \eqref{new action} to that of a coupled BF and Chern--Simons theory.

To continue, let us now focus on the case $\lambda_0\lambda_3\neq\lambda_1\lambda_2$. The time evolution $\partial_t=\lb\H,\cdot\rb$ of the primary constraints is then given by
\be
\partial_t\F(\alpha)\approx\f{1}{2}(\lambda_0\lambda_3-\lambda_1\lambda_2)\M(\alpha),
\q
\partial_t\G(\alpha)\approx\f{1}{2}(\lambda_0\lambda_3-\lambda_1\lambda_2)\N(\alpha),
\ee
where the right-hand side is written in terms of the smeared $3+3$ quantities
\be
\M(\alpha)\equiv\int_\Sigma[\alpha,e_0][\omega\wedge\omega]+2[\alpha,e]\wedge[\omega,\omega_0],
\q
\N(\alpha)\equiv\int_\Sigma[\alpha,\omega_0][e\wedge e]+2[\alpha,\omega]\wedge[e,e_0].
\ee
By projecting onto (or smearing with) $e^i_\mu$ and $\omega^i_\mu$, one can easily show that these quantities are in fact not all independent, but actually satisfy $\M(e_\mu)+\N(\omega_\mu)=0$. This is consistent with the observation made above that $\D$ is the first class constraint generating spacetime diffeomorphisms. For the analysis of the secondary constraints, it is therefore sufficient to focus only on the three components of (say) $\M$. Switching back to a more explicit notation, these are given by
\be
\M^i=2\int_\Sigma\de^2x\,\teps^{ab}\Big(\omega^i_a\big(e^j_0\omega^k_b-\omega^j_0e^k_b\big)-\omega^i_0e^j_a\omega^k_b\Big)\eta_{jk}.
\ee
The stability of the three components of (say) $\F$, which requires the vanishing of $\M$, is therefore equivalent to the two conditions on multipliers
\be
\big(e^i_0\omega^j_a-\omega^i_0e^j_a\big)\eta_{ij}=0,
\ee
and to the single secondary constraint
\be
\S\equiv\teps^{ab}e^i_a\omega^j_b\eta_{ij}\approx0.
\ee
Finally, we now have to check the stability of this secondary constraint. Using $\F$ and $\G$, one can show that its time evolution is given by
\be\nn
\partial_t\S\approx\f{1}{2}e_0\Big(3\lambda_0[e\wedge e]+2\lambda_1[\omega\wedge e]-\lambda_2[\omega\wedge\omega]\Big)+\f{1}{2}\omega_0\Big(\lambda_1[e\wedge e]-2\lambda_2[\omega\wedge e]-3\lambda_3[\omega\wedge\omega]\Big),
\ee
which gives one condition on the Lagrange multipliers. There are therefore no further constraints, and the Dirac algorithm stops here. At the end of the day, we have gathered a total of three conditions on the six multipliers $e^i_0$ and $\omega^i_0$, which leaves three unspecified multipliers corresponding to the three first class constraints (which are the diffeomorphisms).

Out of the six constraints $\F$ and $\G$, we can extract three first class constraints $\D$ corresponding to the diffeomorphisms, while the remaining three constraints will form a second class system together with the secondary constraint $\S$. Putting all this together, the counting therefore shows that there is $(12-(2\times 3+3+1))/2=1$ degree of freedom, as announced.

\section{Perspectives}
\label{sec5}

In this work we have introduced and studied the new non-linear action \eqref{new action} for three-dimensional massive gravity. Although this action is manifestly diffeomorphism-invariant, it does not posses local Lorentz symmetry, and as we have argued in section \ref{sec2} it is precisely this breaking of Lorentz-invariance which is responsible for the appearance of a massive graviton. In addition, this theory is chiral in the sense that is breaks parity invariance (like TMG). At the difference with TMG and NMG however, the new action presented here does not have higher order equations of motion. We have started by giving the simple form of the action in terms of connection and triad variables, which leads to first order equations of motion. Then we have explained how half of the equations of motion can be solved (exactly for $\lambda_3=0$ or perturbatively for $\lambda_3\neq0$) in order to express the connection in terms of the triad. Reinserting this connection in the original action then leads to a pure triad formulation which has second order equations of motion. This second order action (given by \eqref{metric second order l3=0} in the case $\lambda_3=0$) is the sum of a metric contribution, which is the usual Einstein--Hilbert action, and parity breaking terms which can only be expressed in terms of the triad. In this sense, this theory should really be thought of as having the triad as its fundamental dynamical variable.
 
The action \eqref{new action} contains four coupling constants, which are the Planck mass, the cosmological constant, and the two new mass scales coming from the Lorentz-breaking terms. In this sense, it can be thought of as describing a four-parameter family of theories of massive gravity. We have shown that Minkowski spacetime is a solution provided that the coupling constants satisfy a simple algebraic condition. For the linearized perturbations on top of this Minkowski background, we have then found the equations of motion of a model known as self-dual massive gravity, meaning that the linear theory describes a massive graviton. The mass of this graviton depends on the four coupling constants of the new action according to \eqref{generic mass}, and remains non-vanishing when the coupling constants are taken to be vanishing one by one. In the last section of this work, we have studied the full non-linear theory through a detailed Hamiltonian analysis, and shown that it generically propagates the single degree of freedom of a three-dimensional massive graviton. Only when the coupling constants satisfy a simple relation does the theory become topological with no propagating degrees of freedom.

There are many interesting aspects of this new theory which deserve to be studied in more details. First of all, we would like to understand whether its relation with TMG goes beyond linear order (where we have shown the equivalence). If such a relation exists it is potentially very non-trivial, since for example TMG is purely metric and higher order, while the new theory presented here depends (in its triad formulation) on the nine components of the triad and is second order. Furthermore, the TMG action contains a Chern--Simons term for the torsionless Levi--Civita connection, while the theory presented here has non-vanishing torsion. An interesting direction would therefore be to extract the torsionless content of the action \eqref{new action} by imposing the vanishing of the torsion with a Lagrange multiplier in the action. Adding such a term would alter drastically the analysis of this paper, which would have to be repeated, but could lead to interesting results.

Second, it would be very interesting to study further the equations of motion and to analyse whether this theory admis de Sitter or anti-de Sitter spacetime solutions. Can these two other maximally-symmetric spacetimes be solutions of the theory if some conditions between the coupling constants hold, like in the Minkowski case? If not, how are these solutions modified by the new mass terms of the theory? As we have explained at the end of section \ref{sec3.1}, the analysis of the equations of motion and the search for solutions is more subtle than in general relativity because of the breaking of Lorentz symmetry, which forces us to be very careful about the choice of Lorentz frame in which the triad is expressed. Nonetheless, it might be possible to find interesting BTZ-like black hole solutions and to study their stability and thermodynamic properties. One could envision studying aspects of holography \cite{Li:2008dq,Skenderis:2009nt} in this theory, investigating the boundary symmetries along the lines of \cite{Geiller:2017whh,Geiller:2017xad} (and in particular how they are affected by and handle the topological or the various massive limits), or even constructing the quantum theory.

Finally, it would be extremely interesting if the present construction could be extended to four spacetime dimensions, and this direction definitely deserves further investigation. In the first order connection-tetrad formulation, one can also envision preserving diffeomorphism-invariance (at the difference with dRGT) but introducing Lorentz-breaking terms constructed out of contractions of the tetrad $e^I_\mu$ (where $I$ is an internal $\mathfrak{so}(3,1)$ index) and the connection $\omega^{IJ}_\mu$ with the tensors $\eta_{IJ,KL}\equiv\eta_{IK}\eta_{JL}-\eta_{IL}\eta_{JK}$ and $\eps_{IJKL}$. Following what we have done here in three dimensions, one would then consider the terms which are linear in the multipliers $e_0$ and $\omega_0$. The question is then whether it is possible to construct a theory with a non-trivial Minkowski vacuum in which the connection is not vanishing, just like in the case of our equations of motion \eqref{EOMs} (where we have $\bar{\omega}\neq0$), and around which the perturbations could reveal the presence of a massive graviton.

\section*{Acknowledgements}

We would like to thank Jibril Ben Achour, Shinji Mukohyama, Sergey Solodukhin, and Simone Speziale for insightful comments. KN is grateful to the Perimeter Institute for Theoretical Physics for hospitality, where collaboration on this work was initiated.

\appendix

\section{Notations}
\label{appendixNOT}

Throughout this article we denote spacetime indices with Greek letters $\mu,\nu,\ldots\in\{0,1,2\}$, and spatial indices with Latin letters $a,b,\ldots\in\{1,2\}$. Spacetime indices are lowered and raised with the spacetime metric $g_{\mu\nu}$. Latin letters $i,j,\ldots$ from the middle of the alphabet are used to denote $\mathfrak{so}(2,1)$ Lie algebra indices, which are lowered and raised with the internal metric $\eta_{ij}=\text{diag}(-1,1,1)$.

We denote by $\teps_{\mu\nu\rho}=\teps^{\mu\nu\rho}$ the tensor densities of weight $+1$ and $-1$ respectively, which are defined such that $\teps_{012}=1$ in every coordinate system. The spatial restriction of these symbols is denoted by $\teps^{0ab}=\teps^{ab}$. The $\mathfrak{so}(2,1)$ Levi--Civita symbol $\eps_{ijk}$ satisfies the relations
\begin{subequations}
\be
\eps_{ijk}\eps^{lmn}&=-\big(\delta^l_i\delta^m_j\delta^n_k-\delta^l_i\delta^m_k\delta^n_j+\delta^m_i\delta^n_j\delta^l_k-\delta^m_i\delta^n_k\delta^l_j+\delta^n_i\delta^l_j\delta^m_k-\delta^n_i\delta^l_k\delta^m_j\big),\label{epsilon1}\\
\eps_{ijk}\eps^{lmk}&=-\big(\delta^l_i\delta^m_j-\delta^l_j\delta^m_i\big),\label{epsilon2}\\
\eps_{ijk}\eps^{ljk}&=-2\delta^l_i,\\
\eps_{ijk}\eps^{ijk}&=-3!.
\ee
\end{subequations}
With these Levi--Civita symbols we have that the determinant of the triad satisfies
\be\label{determinant}
|e|\equiv\det(e^i_\mu)=-\f{1}{6}\teps^{\mu\nu\rho}\eps_{ijk}e^i_\mu e^j_\nu e^k_\rho,
\q
\teps^{\mu\nu\rho}e^i_\mu e^j_\nu e^k_\rho=|e|\eps^{ijk}.
\ee
Our index-free notation uses the usual definitions and properties of differential forms, such as in particular
\be
e\wedge F=\f{1}{2}\de^3x\,\teps^{\mu\nu\rho}e_\mu F_{\nu\rho},\q e\wedge[e\wedge e]=\de^3x\,\teps^{\mu\nu\rho}e_\mu[e_\nu,e_\rho].
\ee
Furthermore, in this notation there is always an implicit pairing of Lie algebra indices, and $[\cdot\,,\cdot]$ denotes the Lie algebra commutator. We therefore have that
\be
e\wedge F=\eta_{ij}e^i\wedge F^j=e^i\wedge F_i,\q e\wedge[e\wedge e]=\eta_{ij}e^i\wedge[e\wedge e]^j=\eps_{ijk}e^i\wedge e^j\wedge e^k.
\ee
For two one-forms $e$ and $\omega$, and two zero-forms $\alpha$ and $\beta$, we have the useful formulas
\be
[\alpha,e]\wedge[\beta,\omega]+[\alpha,\omega]\wedge[\beta,e]=[\alpha,\beta][e\wedge\omega],
\q
\widetilde{\De}\alpha\wedge\widetilde{\De}\beta=\lambda_2[\alpha,\beta]\widetilde{F},
\ee
while for $(p,q,r)$-forms $(P,Q,R)$ we have
\be
[P\wedge Q]\wedge R=(-1)^{(p+q)r}[R\wedge P]\wedge Q,
\q
[P\wedge Q]=(-1)^{pq+1}[Q\wedge P].
\ee

\section{Topological case}
\label{appendixTOP}

In section \ref{sec3.1} we show that when the condition $\lambda_0\lambda_3=\lambda_1\lambda_2$ holds the theory has six first class constraints and no degrees of freedom. This justifies the name ``topological condition'' for this particular relation between the coupling constants. Here we will show that when this condition is satisfied there is a change of variables which sends the action \eqref{new action} to an action which is indeed manifestly topological.

For this, consider the new variables $\bar{e}$ and $\bar{\omega}$ defined in terms of the initial $e$ and $\omega$ by the invertible change of variables
\be
e=\bar{e}+a\bar{\omega},\q\omega=b\bar{\omega},\q a=\f{\lambda_1}{\lambda_1^2-\lambda_0\lambda_2},\q b=-\f{\lambda_0}{\lambda_1}a=-\f{\lambda_0}{\lambda_1^2-\lambda_0\lambda_2}.
\ee
One can see that this requires that $\lambda_1^2\neq\lambda_0\lambda_2$ as well. Plugging this change of variables in the action \eqref{new action} leads to
\be
S(\bar{e},\bar{\omega})
&=bm_\text{p}\int\bar{e}\wedge\left(\bar{F}+\f{\lambda_0}{6}[\bar{e}\wedge\bar{e}]\right)+a\bar{\omega}\wedge\left(\de\bar{\omega}+\f{1}{3}[\bar{\omega}\wedge\bar{\omega}]\right)+\f{ab}{6\lambda_1}(\lambda_1\lambda_2-\lambda_0\lambda_3)\bar{\omega}\wedge[\bar{\omega}\wedge\bar{\omega}]\nn\\
&=bS_\text{GR}(\bar{e},\bar{\omega})+abS_\text{CS}(\bar{\omega})+\text{``unwanted''}\vphantom{\f{1}{2}}.
\ee
When the topological condition is satisfied, the last term drops, and the action \eqref{new action} is therefore rewritten as the sum of a gravitational action \eqref{action V0} for $\bar{e}$ and $\bar{\omega}$ and a Chern--Simons action for $\bar{\omega}$. This makes manifest the fact that the theory \eqref{new action} has no degrees of freedom when the topological condition is satisfied.

In terms of symmetries, the Hamiltonian analysis has shown that when the topological condition is satisfied there are six first class constraints. These include obviously the diffeomorphisms, but also a hidden Lorentz symmetry, which the above rewriting makes explicit. This Lorentz symmetry can actually be obtained from a certain combination of $\F$ and $\G$ acting on $\bar{e}$ and $\bar{\omega}$. Indeed, the inverse of the above change of variables being given by
\be
\bar{e}=e-\f{a}{b}\omega,
\q
\bar{\omega}=\f{1}{b}\omega,
\ee
one can then use the action of $\F$ and $\G$ on $e$ and $\omega$ to compute
\be
\lb a\F(\alpha)+b\G(\alpha),\bar{e}\rb=[\bar{e},\alpha],
\q
\lb a\F(\alpha)+b\G(\alpha),\bar{\omega}\rb=\de\alpha+[\bar{\omega},\alpha],
\ee
which shows as expected that the combination $a\F+b\G$ generates infinitesimal Lorentz transformations of the new variables $\bar{e}$ and $\bar{\omega}$.

\section{Relation to zwei-Dreibein gravity}
\label{appendixBI}

In this appendix we explain for the sake of completeness the relationship between our new action \eqref{new action} and the three-dimensional bi-metric theory known as zwei-Dreibein gravity. Let us take as the starting point the zwei-Dreibein action of \cite{Alexandrov:2014oda} given by equation (2.1). This action depends on two triad fields and two connections, and can be written as
\be
S(e_+,e_-,\omega_+,\omega_-)&=\int e_+\wedge F_++e_-\wedge F_--\f{\Lambda_+}{6}e_+\wedge[e_+\wedge e_+]-\f{\Lambda_-}{6}e_-\wedge[e_-\wedge e_-]\nn\\
&\phantom{\ =\int e_+\wedge F_++e_-\wedge F_-}-\f{\beta_+}{2}e_-\wedge[e_+\wedge e_+]-\f{\beta_-}{2}e_+\wedge[e_-\wedge e_-].
\ee
On the first line we recognize the sum of two gravitational actions \eqref{action V0} with a cosmological constant, and on the second line are two coupling terms between the triads (or the two metrics) $e_+$ and $e_-$. With the choice
\be
e_+=e,\q e_-=\omega,\q\omega_+=\omega,\q\omega_-=0,
\ee
this action then becomes
\be
S(e,\omega)=\int e\wedge F-\f{\Lambda_+}{6}e\wedge[e\wedge e]-\f{\Lambda_-}{6}\omega\wedge[\omega\wedge\omega]-\f{\beta_+}{2}\omega\wedge[e\wedge e]-\f{\beta_-}{2}e\wedge[\omega\wedge\omega],
\ee
and one finally obtains \eqref{new action} by setting the coupling constants to
\be
\Lambda_+=-\lambda_0,\q\beta_+=-\lambda_1,\q\beta_-=1-\lambda_2,\q\Lambda_-=-\lambda_3,
\ee
and rescaling by an overall factor of $m_\text{p}$.

Notice that this map, because it involves setting one of the two initial connections to zero, is of course not an innocent invertible change of variables. This is indeed to be expected since the three-dimensional bi-metric theories actually propagate two degrees of freedom instead of one, and therefore describe very different physics from the new action \eqref{new action}. Obviously, identifying the two triads and the two connections in the zwei-Dreibein action leads to the action for general relativity, and not to a theory with a single degree of freedom. The theory \eqref{new action} can therefore in a sense be thought of as living ``in between'' general relativity and the zwei-Dreibein theory: it has a single set of gravitational data, i.e. a triad and a connection, but still propagates one degree of freedom. Moreover, it is interesting to note that the Hamiltonian analysis of section \ref{sec4} is very similar in spirit to that of \cite{Alexandrov:2014oda}, and that in this reference the authors have carefully studied the issue of partial masslessness of one of the two degrees of freedom, which is also a bit reminiscent of the topological and massless limits which exist for \eqref{new action}.

\section{Extra kinetic terms}
\label{appendixKIN}

In this appendix we discuss the possibility of having other kinetic terms in the action \eqref{new action}, and the condition under which they can be eliminated by a change of variables. Indeed, one could in principle consider the most general (first order) kinetic terms constructed out of $e$ and $\omega$, and study the action
\be\label{generic kinetics}
S(e,\omega)=m_\text{p}\int\alpha_1e\wedge\de\omega+\f{\alpha_2}{2}e\wedge\de e+\f{\alpha_3}{2}\omega\wedge\de\omega+V(e,\omega),
\ee
where the potential is again \eqref{general potential}.

Given this action, it is natural to ask whether there can exist a change of variables which eliminates the kinetic terms in $\alpha_2$ and $\alpha_3$. If this is possible, then we can conclude that we can take $\alpha_2=\alpha_3=0$ without loss of generality. To investigate this, consider the new variables $\bar{e}$ and $\bar{\omega}$ defined in terms of the initial $e$ and $\omega$ by the invertible change of variables
\be
e=a\bar{e}+b\bar{\omega},
\q
\omega=c\bar{e}+d\bar{\omega},
\q
ad-bc\neq0.
\ee
The nice property of the potential \eqref{general potential} is that it already contains the four possible terms which can be constructed out of $e$ and $\omega$, and will therefore have the same form when expressed in terms of the new variables, but simply contain new coupling constants $\bar{\lambda}_n(\lambda_n,a,b,c,d)$. To know whether the action \eqref{generic kinetics} can be rewritten in the form \eqref{new action}, it is thus sufficient to focus on the fate of the kinetic terms.

Obviously, the action expressed with the new variables will also contain the three possible kinetic terms, but now with new coupling constants given by
\be
\bar{\alpha}_1=ad\big[(1+xy)\alpha_1+y\alpha_2+x\alpha_3\big],
\quad
\bar{\alpha}_2=a^2(\alpha_3x^2+2\alpha_1x+\alpha_2),
\quad
\bar{\alpha}_3=d^2(\alpha_2y^2+2\alpha_1y+\alpha_3),
\ee
where we have introduced $x\equiv c/a$ and $y\equiv b/d$. If the condition $\Delta\equiv\alpha_1^2-\alpha_2\alpha_3>0$ is satisfied, it is therefore always possible to find real coefficients $(a,b,c,d)$ which can set $\bar{\alpha}_2=\bar{\alpha}_3=0$ while satisfying $ad-bc\neq0$. For this we simply need to choose
\be
x=\f{-\alpha_1\pm\sqrt{\Delta}}{\alpha_3},
\q
y=\f{-\alpha_1\pm\sqrt{\Delta}}{\alpha_2},
\ee
with the same sign $\pm$ in both solutions in order to have $ad-bc\neq0$. If however we have $\Delta<0$, then the transformation which eliminates the kinetic terms in $\bar{\alpha}_2$ and $\bar{\alpha}_3$ still exists but becomes imaginary, which could indicate that the theory might then propagate more than one degree of freedom. This is an interesting point which should be studied with more care.

\section{Matrix formulation}
\label{appendixMAT}

We explain in this appendix how to rewrite the action \eqref{new action} in terms of the two $3\times3$ matrices
\be
\Omega^{ij}\equiv\teps^{\mu\nu\rho}e^i_\mu\partial_\nu e^j_\rho,
\q
M^{ij}\equiv\omega^i_\mu\hat{e}^{\mu j},
\ee
where $\hat{e}$ is the inverse of $e$ in the sense that $e^i_\mu\hat{e}^\mu_j=\delta^i_j$ and $e^i_\mu\hat{e}^\nu_i=\delta^\nu_\mu$. To obtain this rewriting, we proceed by analyzing and rewriting all the terms in the first line of \eqref{new action indices}.

First, since $M^{ij}=\omega^i_\mu\hat{e}^{\mu j}$ and $\hat{e}^{\mu j}e_{\rho j}=\delta^\mu_\rho$, we have $\omega^i_\rho=M^{ij}e_{\rho j}$. Up to a total derivative obtained from the integration by parts, we can therefore rewrite the kinetic term in \eqref{new action indices} as
\be
\teps^{\mu\nu\rho}e_{\mu i}\partial_\nu\omega^i_\rho=-\teps^{\mu\nu\rho}\partial_\nu e_{\mu i}\omega^i_\rho=-\teps^{\mu\nu\rho}\partial_\nu e_\mu^iM_{ij}e_\rho^j=\teps^{\mu\nu\rho}e_\mu^j\partial_\nu e_\rho^iM_{ij}=\Omega^{ji}M_{ij}=\tr(\Omega M).
\ee
Then, for the term in $\lambda_0$, with the definition \eqref{determinant} of the determinant of the triad we have that
\be
\f{1}{6}\teps^{\mu\nu\rho}\eps_{ijk}e^i_\mu e^j_\nu e^k_\rho=-|e|.
\ee
For the term in $\lambda_1$, we can use again \eqref{determinant} to write that
\be
\f{1}{2}\teps^{\mu\nu\rho}\eps_{ijk}\omega^i_\mu e^j_\nu e^k_\rho=\f{1}{2}\teps^{\mu\nu\rho}\eps_{ijk}M^{il}e_{\mu l}e^j_\nu e^k_\rho=\f{|e|}{2}\eps_{ijk}{\eps_l}^{jk}M^{il}=-|e|{M^i}_i=-|e|\tr(M).
\ee
Using the same manipulation, we then get for the term in $\lambda_2$ that
\be
\f{1}{2}\teps^{\mu\nu\rho}\eps_{ijk}\omega^i_\mu\omega^j_\nu e^k_\rho=\f{1}{2}\teps^{\mu\nu\rho}\eps_{ijk}M^{il}M^{jm}e_{\mu l}e_{\nu m}e^k_\rho=\f{|e|}{2}\eps_{ijk}{\eps_{lm}}^kM^{il}M^{jm},
\ee
and expanding the product of Levi--Civita symbols using \eqref{epsilon2} then leads to
\be
\f{1}{2}\teps^{\mu\nu\rho}\eps_{ijk}\omega^i_\mu\omega^j_\nu e^k_\rho=\f{|e|}{2}({M^i}_j{M^j}_i-{M^i}_i{M^j}_j)=\f{|e|}{2}\big[\tr(M^2)-\tr^2(M)\big],
\ee
with an obvious notation for the product of traces and the trace of a matrix product. Finally, using the definition of the matrix determinant as
\be
\det(M)=-\f{1}{6}\eps_{ijk}\eps^{lmn}{M^i}_l{M^j}_m{M^k}_n,
\ee
we have
\be
\f{1}{6}\teps^{\mu\nu\rho}\eps_{ijk}\omega^i_\mu\omega^j_\nu\omega^k_\rho=\f{1}{6}\teps^{\mu\nu\rho}\eps_{ijk}M^{il}M^{jm}M^{kn}e_{\mu l}e_{\nu m}e_{\rho n}=-|e|\det(M),
\ee
and using formula \eqref{epsilon1} then enables us to write
\be
\det(M)
&=\f{1}{6}\big({M^i}_i{M^j}_j{M^k}_k-{M^i}_i{M^j}_k{M^k}_j+{M^i}_k{M^j}_i{M^k}_j\nn\\
&\phantom{\ =-\f{1}{6}\big(}-{M^i}_j{M^j}_i{M^k}_k+{M^i}_j{M^j}_k{M^k}_i-{M^i}_k{M^j}_j{M^k}_i\big)\nn\\
&=\f{1}{6}\big[\tr^3(M)+2\tr(M^3)-3\tr(M)\tr(M^2)\big].
\ee
Putting these ingredients together gives the form \eqref{matrix action} of the action.

\section{Solution for the connection}
\label{appendixSOL}

In this appendix we give the various equivalent expressions for the solution of the equations of motion \eqref{eom omega} giving $\omega$ as a function of $e$ in the case $\lambda_3=0$. With the matrix notation, we have obtained the solution \eqref{solution for M} for $M$ in terms of $\Omega$. Using the fact that $\omega^i_\mu=M^{ij}e_{\mu j}$ together with formula \eqref{determinant} leads to the explicit expression
\be\label{explicit solution omega}
\omega^i_\mu
=\f{1}{\lambda_2|e|}\left(\f{1}{2}{\Omega^j}_je_\mu^i-\Omega^{ij}e_{\mu j}\right)-\f{\lambda_1}{2\lambda_2}e^i_\mu
=\f{1}{2\lambda_2}\big(e^i_\mu{\eps_j}^{kl}-2\eps^{ikl}e_{\mu j}\big)\partial_\nu e_\rho^j\hat{e}^\nu_k\hat{e}^\rho_l-\f{\lambda_1}{2\lambda_2}e^i_\mu.
\ee
As a consistency check, one can verify that this is in agreement with the ``usual'' way of solving the equations of motion \eqref{eom omega} for $\omega$. This requires using the explicit inversion formula which gives $\omega$ in terms of $W$ and $e$ whenever $[\omega\wedge e]=W$ for some Lie algebra-valued two-form $W$, namely
\be
{\eps^i}_{jk}\big(\omega^j_\mu e^k_\nu-\omega^j_\nu e^k_\mu\big)=W^i_{\mu\nu}
\q\Leftrightarrow\q
\omega^i_\mu=\f{1}{4}e^i_\mu{\eps_j}^{kl}W^j_{\nu\rho}\hat{e}^\nu_k\hat{e}^\rho_l+{\eps^i}_{jk}W^j_{\mu\nu}\hat{e}^{\nu k}.
\ee
Noting that the equations of motion \eqref{eom omega} with $\lambda_3=0$ take the explicit form
\be
{\eps^i}_{jk}\big(\omega^j_\mu e^k_\nu-\omega^j_\nu e^k_\mu\big)=-\f{1}{\lambda_2}\big(\partial_\mu e^i_\nu-\partial_\nu e^i_\mu+\lambda_1{\eps^i}_{jk}e^j_\mu e^k_\rho\big),
\ee
we get the solution
\be
\omega^i_\mu=-\f{1}{2\lambda_2}\Big(e^i_\mu{\eps_j}^{kl}\partial_\nu e^j_\rho\hat{e}^\nu_k\hat{e}^\rho_l+2\eps^{ijk}\big(\partial_\mu e_{\nu j}-\partial_\nu e_{\mu j}\big)\hat{e}^\nu_k\Big)-\f{\lambda_1}{2\lambda_2}e^i_\mu.
\ee
Although this looks actually different from \eqref{explicit solution omega}, these two expressions can be shown to be identical upon computing a double anti-symmetrization of the internal indices, and in turn equal to
\be\label{torsion-free connection}
\omega^i_\mu=-\f{1}{2}\eps^{ijk}\eps_{jkl}\omega^l_\mu=\f{1}{2\lambda_2}{\eps^i}_{jk}\hat{e}^{\nu j}\big(\partial_\mu e_\nu^k-\partial_\nu e_\mu^k-\hat{e}^{\rho k}e^l_\mu\partial_\nu e_{\rho l}\big)-\f{\lambda_1}{2\lambda_2}e^i_\mu=\f{1}{2\lambda_2}{\eps^i}_{jk}\Gamma^{jk}_\mu-\f{\lambda_1}{2\lambda_2}e^i_\mu,
\ee
where we can now recognize when $\lambda_1=0$ and $\lambda_2=1$ the more familiar expression for the torsion-free connection $\omega$ in terms of the Levi--Civita connection $\Gamma^{jk}_\mu=\hat{e}^{\nu j}\nabla_\mu e^k_\nu=\hat{e}^{\nu j}(\partial_\mu e^k_\nu-\Gamma^\rho_{\mu\nu}e^k_\rho)$.

\section{Triad formulation for small $\boldsymbol{\lambda_3}$}
\label{appendix6}

In this appendix we study how the pure triad action \eqref{second order l3=0} is modified when we include a small but non-vanishing value of the coupling constant $\lambda_3$. For this, let us go back to the matrix form \eqref{matrix action} of the action, and write the equations of motion for $M$ when $\lambda_3\neq0$. These are given by
\be
\Omega-|e|\Big(\lambda_1\eta+\lambda_2\big[\tr(M)\eta-M\big]+\lambda_3\,\text{det}'(M)\Big)=0,
\ee
where
\be
\text{det}'(M)=\f{1}{2}\big[\tr(M^2)-\tr^2(M)\big]\eta+M\big[\tr(M)\eta-M\big].
\ee
Denoting by $M_0$ the solution \eqref{solution for M} obtained for $\lambda_3=0$, i.e. the matrix such that
\be\label{eom omega M0}
\Omega-|e|\Big(\lambda_1\eta+\lambda_2\big[\tr(M_0)\eta-M_0\big]\Big)=0,
\ee
we look for first order corrections of the form $M=M_0+\lambda_3M_1$. Plugging this ansatz in the equations of motion, using \eqref{eom omega M0} and then keeping only the terms of order $\lambda_3$ leaves us with the equation
\be
\lambda_2\big[\tr(M_1)\eta-M_1\big]+\text{det}'(M_0)=0.
\ee
This equation can then obviously be solved to find $M_1$ in terms of $M_0$, which is therefore an expression for $M_1$ in terms of $\Omega$. This explicit solution is however rather lengthy and in fact not necessary for our purposes.

Indeed, to see how we can completely bypass this more complicated calculation, let us simply plug the ansatz $M=M_0+\lambda_3M_1$ in the action \eqref{matrix action}. Keeping only the terms of order $\lambda_3$ leads to
\be
S(e)=S_0(e)+\lambda_3S_1(e),
\ee
where $S_0(e)$ has been computed in \eqref{second order l3=0} and the first order correction is
\be
S_1(e)=m_\text{p}\int\de^3x\,\Big\{\tr(\Omega M_1)-|e|\Big(\lambda_1\tr(M_1)+\lambda_2\big[\tr(M_0)\tr(M_1)-\tr(M_0M_1)\big]+\det(M_0)\Big)\Big\}.
\ee
Now, multiplying \eqref{eom omega M0} by $M_1$ and taking the trace of the resulting equation leads to the identity
\be
\tr(\Omega M_1)-|e|\Big(\lambda_1\tr(M_1)+\lambda_2\big[\tr(M_0)\tr(M_1)-\tr(M_0M_1)\big]\Big)=0.
\ee
This then dramatically simplifies the expression for the first order correction to the action, which can be expressed solely in terms of $M_0$ and becomes
\be
S_1(e)=-m_\text{p}\int\de^3x\,|e|\det(M_0)=-\f{m_\text{p}}{6}\int\de^3x\,|e|\big[\tr^3(M_0)+2\tr(M_0^3)-3\tr(M_0)\tr(M_0^2)\big].
\ee
Using the solution \eqref{solution for M} for $M_0$, we can then compute the explicit expression
\be\label{action S1}
S_1(e)
&=\f{m_\text{p}}{2\lambda_2}\int\de^3x\left\{\f{\lambda_1}{4\lambda_2^2|e|}\left[\tr^2(\Omega)-2\tr(\Omega^2)\right]+\f{\lambda_1^3}{4\lambda_2^2}|e|-\f{\lambda_1^2}{4\lambda_2^2}\tr(\Omega)\right.\nn\\
&\phantom{\ =\f{m_\text{p}}{2\lambda_2}\int\de^3x\left\{\right.}\left.+\f{1}{12\lambda_2^2|e|^2}\left[\tr^3(\Omega)-6\tr(\Omega)\tr(\Omega^2)+8\tr(\Omega^3)\right]\right\}.
\ee
Putting the two contributions \eqref{second order l3=0} and \eqref{action S1} together, we finally get that the triad action to first order in $\lambda_3$ takes the form
\be\label{second order l3 small}
S(e)
&=\f{m_\text{p}}{2\lambda_2}\int\de^3x\left\{\f{\mu_2}{2|e|}\big[\tr^2(\Omega)-2\tr(\Omega^2)\big]+\f{\mu_0}{2}|e|-\mu_1\tr(\Omega)\right.\nn\\
&\phantom{\ =\f{m_\text{p}}{2\lambda_2}\int\de^3x\left\{\right.}+\left.\f{\mu_3}{|e|^2}\big[\tr^3(\Omega)-6\tr(\Omega)\tr(\Omega^2)+8\tr(\Omega^3)\big]\right\},
\ee
with
\be
\mu_0\equiv3\lambda_1^2-4\lambda_0\lambda_2+\f{\lambda_1^3\lambda_3}{2\lambda_2^2},
\q
\mu_1\equiv\lambda_1+\f{\lambda_1^2\lambda_3}{4\lambda_2^2},
\q
\mu_2\equiv1+\f{\lambda_1\lambda_3}{2 \lambda_2^2},
\q
\mu_3\equiv\f{\lambda_3}{12\lambda_2^2}.
\ee
The triad action at first order in $\lambda_3$ is therefore given by the Einstein--Hilbert action with a cosmological constant (these are the terms in $\mu_2$ and $\mu_0$), augmented by two ``massive modifications'', which are the terms in $\mu_1$ and $\mu_3$. The first of these modifications is of course already present in \eqref{second order l3=0}, while the term in $\mu_3$ appears because $\lambda_3\neq0$.

Finally, we can now proceed to a consistency check and show that the action \eqref{second order l3 small}, when perturbed around a Minkowski background, reproduces a graviton mass consistent with the general result of section \eqref{sec3.4}. From the expressions \eqref{linear Omega}, one can see that the term in $\mu_3$ in \eqref{second order l3 small} will not contribute to the action for the perturbations since it will be cubic in $f$. Therefore, the calculation reduces to that of section \eqref{sec3.3} where we simply replace $\lambda_1\rightarrow\mu_1/\mu_2$ and $\lambda_2\rightarrow\lambda_2/\mu_2$. In particular, this means that starting from \eqref{second order l3 small} the graviton mass will be given by
\be
\f{\mu_1}{\mu_2}\simeq\lambda_1-\f{\lambda_1^2\lambda_3}{4\lambda_2^2},
\ee
where we have kept the lowest order in $\lambda_3$. One can then see that this result is indeed consistent with the generic graviton mass \eqref{generic mass} when it is approximated at lowest order in $\lambda_3$, i.e.
\be
m_\text{g}\simeq\lambda_1+\left(\lambda_0-\f{\lambda_1^2}{\lambda_2}\right)\f{\lambda_3}{\lambda_2}\simeq\lambda_1-\f{\lambda_1^2\lambda_3}{4\lambda_2^2},
\ee
where for the second equality we have used the condition $\mu_0=0$. This condition is simply the fact that the cosmological constant has to vanish in order for the Minkowski background to be a solution of the triad action at first order in $\lambda_3$.

\bibliography{3Dmassive_biblio}
\bibliographystyle{Biblio}

\end{document}